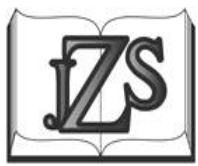

# Effect of IBA concentration and water soaking on rooting hardwood cuttings of black mulberry (*Morus nigra* L.)

Rasul Rafiq Aziz[1], Aram Akram Mohammed[1*], Faraydwn Karim Ahmad[1] & Ari Jamil Ali[2]

[1] Horticulture Department, College of Agricultural Sciences, Engineering University of Sulaimani, Sulaimani, Kurdistan Region-Iraq.
[2] Horticulture and Landscape Design, Halabja Technical College of Applied Sciences, Sulaimani Polytechnic University, Kurdistan Region-Iraq.
*Corresponding author's e-mail: aram.hamarashed@univsul.edu.iq*

| Article info | Abstract |
|---|---|
|  | The research was conducted at the College of Agricultural Sciences Engineering/University of Sulaimani/ Kurdistan Region-Iraqi to investigate effects of different concentrations of IBA (0, 3000, 4000 and 5000 ppm) and soaking in water for 24 hours on propagation black mulberry (*Morus nigra* L.) by hardwood cuttings. In this research the parameters of rooting percentage, root number, root length, sprout bud number, shoot length and shoot diameter were measured. Effect of individual factors showed that the highest rooting percentage (15%) was achieved in cuttings soaked in water for 24 hours, as well as improving other traits. Also, the best (23.33%) rooting was found in cuttings dipped in 4000 ppm IBA. Interaction effects of the two factors showed that cuttings treated with 4000 ppm IBA and soaked in water for 24 hours gave the highest (40%) rooting, and the highest other root and shoot traits were achieved in the same interaction as well. |

## Introduction

Mulberry (*Morus sp.*) is a deciduous fruit which attributed to *moraceae* family, and it has been adapted to a wide range of tropical, subtropical, and temperate zones [13]. The common methods are used in propagation of mulberry species are cuttings, layering, and budding. Despite, the main conditions in propagation plants, particularly in fruit trees, are reducing costs, true-to-type and obtain large numbers of new plants simultaneously. These reasons encourage propagators to use feasible methods. The most feasible and cost-effective method is stem cuttings propagation. In connection with these points, [6] referred that cutting propagation is three times cheaper than grafting. Also, stem cuttings are beneficial to obtain large numbers of new plants, even from one stock plant with the same genetic property as parent plant [17]. One of the least expensive and easiest type of stem cuttings is hardwood cuttings, however many external and internal factors affect the rooting of cuttings such as growth regulators [7]. Indole-3-butyric acid (IBA) is a growth regulator from auxin group induces rooting in different types of stem cuttings due to it is stable, less toxic and insensitive to the auxin degrading enzymes compared with NAA and 2,4-D [12].

Indole-3-butyric acid (IBA) is essential for root initiation in hardwood cuttings of black mulberry, which is a difficult-to-root fruit species for most cultivars, whereas there is no optimal concentration of IBA for rooting black mulberry hardwood cuttings. [16] noticed that hardwood cuttings of mulberry cultivars could not root without indole-3-butyric acid (IBA), and 5000 ppm was the best dose for rooting, while [10] summarized that 4000 ppm IBA was the best, [4] observed 8000 ppm IBA was the best dose for rooting hardwood cuttings of black mulberry. Another factor may affect stem cuttings to produce new plant is soaking in water. Soaking cuttings in water for a period improves rooting by leaching out the rooting inhibitors and has effects on many other biochemical aspects in cuttings [15]. In this regard, the aim of this research to determine the best concentration of IBA and whether soaking in water has effect on rooting hardwood cuttings of black mulberry or not.

**Materials and methods**

The experiment was carried out in the dormant season, at the College of Agricultural Sciences Engineering/University of Sulaimani/Iraqi-Kurdistan- region from 1 December 2018 to 28 may 2019. Hardwood cuttings of black mulberry (*Morus nigra* L.) were collected on 1 December 2018 from a local cultivar grown in Kurdistan region which known as (Shatu), and cuttings prepared from the basal part of one-year-old shoots of 10-year-old trees.

Immediately after the collection, the cuttings were cut into 20 cm long and 10-13 mm diameter, after that the basal portion of 60 cuttings up to 1.5-2 cm were dipped for 10 seconds in different concentrations of IBA (0, 3000, 4000 and 5000) ppm, which dissolved in 50% ethanol. At the time of planting, 2/3 part of cutting was inserted in sand medium in polyethylene bags.

On the other hand, another 60 cuttings were soaked in water for 24 hours, the following day they taken out from the water and after drained the free water treated with the same IBA concentrations respectively, and then planted in sand media which placed in polyethylene bags with size 15×30 cm, and 5 cuttings stuck in each polyethylene bag, 3 replications used for each treatment.

The experiment was laid down in complete randomized block design (RCBD) in an uncontrolled greenhouse. On 28 May 2019 the experiment was terminated and the parameters of rooting percentage, number of main roots, length of main roots, sprout bud number, shoot length and shoot diameter were measured. The data were analyzed by computer program XLSTAT, Duncun's multiple range test (5%) was used for comparisons of means.

**Result and discussion**

The results in table (1) shows that effect of water soaking treatment on rooting percentage of black mulberry hardwood cuttings was significant compared with the cuttings were not soaked in the water. The highest (15%) rooting percentage was achieved in cuttings soaked in water for 24 hours, while the cuttings were not soaked gave (8.33%) rooting percentage. Also, soaking cuttings in water for 24 hours improved root length significantly in comparison with cuttings were not soaked in water, the longest root (3.35 cm) was observed in cuttings soaked in water for 24 hours, but the cuttings were not soaked gave the shortest root (1.16 cm). However, root number, sprout bud number, shoot length and shoot diameter were not significantly different in cuttings with soaking for 24 hours and in those cuttings without soaking. Increasing root percentage in soaked cuttings may be ascribed to that soaking in water increased water content of the cuttings, and managing water content of cuttings, particularly at the beginning of planting cuttings, is crucial for surviving cuttings, because cuttings at that time are rootless and have no callus, so losing water from cuttings are very difficult to compensate, consequently the majority of cuttings will die. [19] reported that soaking cuttings in water pre-planting enhances water content of the cuttings and avoids cuttings from water stress and desiccation during early stage of planting of cuttings. In addition to, soaking cuttings in water may remove the substances that inhibit rooting and increase the substances that promote rooting. [2] and [3] described that soaking cuttings in water causes removing of substances inhibit rooting and stimulates ethylene production. Moreover, all species of mulberry contain latex [8], latex may have adverse effect on rooting, and soaking in water leaches the latex, particularly at the cutting bases, before dipping in IBA, finally this may improve rooting. [6] referred that soaking cuttings of *Sciadopitys verticillata* for 24 hours in water before IBA application enhanced rooting as a result of removing latex sap at the bases of the cuttings.

**Table 1:** Effect of water soaking treatment on root and shoot characteristics of hardwood cuttings of black mulberry.

| Water soaking treatment | Rooting % | Main root number | Main root length (cm) | Sprout bud number | Shoot length (cm) | Shoot diameter (mm) |
|---|---|---|---|---|---|---|
| Without soaking | 8.33 b | 0.83 a | 1.16 b | 0.58 a | 0.83 a | 1.34 a |
| With soaking | 15 a | 1.45 a | 3.35 a | 0.76 a | 1.05 a | 1.60 a |

* The values in each column with the same letter do not differ significantly ($P \leq 0.05$) according to Duncan's Multiple Range Test.

The results of the effects of different IBA concentrations on hardwood cuttings of black mulberry shown in table (2) demonstrated that IBA concentrations were effective for rooting black mulberry hardwood cuttings compared to control cuttings. Cuttings dipped in 4000 ppm IBA significantly different with control cuttings with and without soaking in water. The concentration of 4000 ppm IBA resulted the highest (23.3%) rooting percentage, while no rooting observed in control cuttings in the both case with and without soaking in water, however rooting decreased by increasing IBA concentration to 5000 ppm. Correspondingly, [10] found 4000 ppm IBA as the best for rooting of black mulberry hardwood cuttings. Auxins, particularly IBA, induce adventitious root formation in cuttings by different ways. The histological studies revealed that IBA application induce root formation by increasing cell division at cambial zone from which root primordia finally arise [18]. Besides, it was considered that application IBA to cuttings exogenously increased endogenous IAA by converting to IAA, and IBA may improve IAA action or synthesis of IAA in cuttings, also IBA can enhance tissue sensitivity for IAA and increase rooting [20, 22]. Moreover, the IBA used to the bases of cuttings translocated greater than IAA to the upper end, and quickly metabolized into IBA conjugates which play role as auxin sources in the following stages of rooting [21]. In spite of these, an increase in metabolic activity was observed in *Dalbergia sissoo* leafy cuttings because of application IBA. Carbohydrate metabolism may have activated by IBA to obtain energy, and in the rooting zone of cuttings during adventitious root primordium initiation and development, protein and PEP-activity needed for cell division and differentiation [9].

Besides, effect of the 4000 IBA concentration on root length and shoot diameter was significant as well, the longest (6.08 cm) root and the thickest (2.88 mm) shoot diameter were found in 4000 ppm IBA, but control cuttings gave no results. These are may be due to 4000 ppm IBA give rise to earlier rooting and cell division, and also the roots might have more time to grow and absorb nutrients by which shoot growth might be improved. [14] postulated that improving root length and shoot growth at 2000 ppm IBA in peach cuttings might be due to it promoted earlier rooting and quick cell division. The three IBA concentrations and control showed no significant results in root number, sprout bud number and shoot length.

**Table 2:** Effect of IBA concentrations on root and shoot characteristics of hardwood cuttings of black mulberry.

| IBA concentration (ppm) | Rooting % | Main root number | Main root length (cm) | Sprout bud number | Shoot length (cm) | Shoot diameter (mm) |
|---|---|---|---|---|---|---|
| Control | 0 b | 0 a | 0 b | 0 a | 0 a | 0 b |
| 3000 | 10 ab | 1.25 a | 1.62 b | 0.66 a | 0.54 a | 1.20 ab |
| 4000 | 23.33 a | 2.16 a | 6.08 a | 1.2 a | 1.72 a | 2.88 a |
| 5000 | 13.33 ab | 1.16 a | 1.33 b | 0.83 a | 1.50 a | 1.79 ab |

* The values in each column with the same letter do not differ significantly ($P \leq 0.05$) according to Duncan's Multiple Range Test.

Interaction effects of soaking in water and IBA concentrations showed in table (3) revealed that rooting percentage in cuttings soaked in water for 24 hours and treated with 4000 ppm IBA produced significant difference with control cuttings with and without soaking. The highest (40%) rooting percentage was observed in cuttings treated with 4000 ppm IBA and soaked in water, while the control cuttings with and without soaking gave no rooting and other results. At the same time, interaction of 4000 ppm IBA with soaking in water gave the best (3.66) root number, (9.83 cm) root length, (1.73) sprout bud number, (2.45 cm) shoot length and (4.14 mm) shoot diameter. In a similar way, [5] summarized that soaking hardwood cuttings of 41B grapevine rootstock in water at least for 24 hours and treated with IBA afterwards gave the best results, since soaking in water may leach rooting inhibitors from cuttings and addition of IBA may provide further benefits. Regarding, [11] and [1] found that soaking cuttings of grapevine rootstock 140 Ruggeri in water decreased GA-like compounds which inhibits rooting, and raised IAA level that promotes rooting, hence rooting ability was enhanced.

**Table 3**: Interaction effects of soaking water treatment and IBA concentrations on root and shoot characteristics of hardwood cuttings of black mulberry.

| Water soaking treatment | IBA concentration (ppm) | Rooting % | Main root number | Main root length (cm) | Sprout bud number | Shoot length (cm) | Shoot diameter (mm) |
|---|---|---|---|---|---|---|---|
| Without soaking | Control | 0 b | 0 b | 0 b | 0 b | 0 b | 0 b |
| | 3000 | 6.66 b | 1 ab | 0.33 b | 0.66 ab | 0.66 ab | 1.16 b |
| | 4000 | 6.66 b | 0.66 b | 2.33 b | 0.66 ab | 1 ab | 1.62 ab |
| | 5000 | 20 ab | 1.66 ab | 2 b | 1 ab | 1.66 ab | 2.58 ab |
| With soaking | Control | 0 b | 0 b | 0 b | 0 b | 0 b | 0 b |
| | 3000 | 13.33 ab | 1.5 ab | 2.91 b | 0.66 ab | 0.41 ab | 1.25 ab |
| | 4000 | 40 a | 3.66 a | 9.83 a | 1.73 a | 2.45 a | 4.14 a |
| | 5000 | 6.66 b | 0.66 b | 0.66 b | 0.66 ab | 1.33 ab | 1 b |

\* The values in each column with the same letter do not differ significantly ($P \leq 0.05$) according to Duncan's Multiple Range Test.

**Conclusion**

In conclusion, hardwood cuttings of black mulberry were soaked in water for 24 hours gave the best rooting percentage and root length. 4000 ppm IBA was the best dose for rooting of the cuttings, and also gave the best root length and shoot diameter. Effect of soaking in water and IBA concentration on hardwood cuttings of black mulberry were the best when they interacted together. Interaction of soaking in water for 24 hours and 4000 ppm IBA enhanced all studied parameters.

**References**

[1] Bartolini, G., Toponi, M. A. and L. *"Santini. Endogenous GA-like substances in dipping waters of cuttings of two Vitis rootstocks"*. American journal of enology and viticulture, Vol. 37, No. 1, pp. 1-6. (1986).

[2] Blake, T. J., Cain, N. P., Gadziola, R., Hills, C.I. and T. Zitnak. *"Ethylene and auxin improve the rooting of difficult-to-root poplar cuttings. Proc. North American Poplar Council Annual Meeting"*. Rhinelander, WI, pp. 20–28. (1982).

[3] DesRochers, A. and B. R. Thomas. *"A comparison of pre-planting treatments on hardwood cuttings of four hybrid poplar clones"*. New Forests, Vol. 26, No. 1, pp. 17-32. (2003).

[4] Edİzer, Y., Gökçek, O., and O. Saraçoğlu. "Effects of growth regulators application on propagation with hardwood cuttings of the black mulberry". Gaziosmanpaşa Üniversitesi Ziraat Fakültesi Dergisi, Vol. 33, No. 3, pp. 92-96. (2016).

[5] Gökbayrak, Z., Dardeniz, A., Arıkan, A. and U. Kaplan. *"Best duration for submersion of grapevine cuttings of rootstock 41B in water to increase root formation"*. Journal of Food, Agriculture & Environment, Vol. 8, No. 3-4, pp. 607-609. (2010).

[6] Hartmann, H.T. Kester, D.E. Davies Júnior, F.T. and R.L. Geneve. *"Plant propagation: principles and practices"*. 8th ed. New Jersey: Prentice Hall, p. 915. (2011).

[7] Hartmann, H.T. Kester, D.E. Davies, F.T. and R.L. Geneve. *"Plant Propagation: Principles and Practices"*. (7th edition), New Jersey. (2002).

[8] Hora, B. *"The Oxford encyclopedia of trees of the world"*. Oxford, UK: Oxford University Press. 288 p. (1981).

[9] Husen, A. *"Clonal propagation of Dalbergia sissoo Roxb. and associated metabolic changes during adventitious root primordium development"*. New Forests, Vol.36, No.1, pp.13-27. (2008).

[10] Kako, S.M. *"The effect of auxin IBA and Kinetin in budding success percentage of mulberry (Morus sp.)"*. International Journal of Pure and Applied Sciences and Technology, Vol.13, No. 1, pp. 50-56. (2012).

[11] Kracke, H., Cristoferi, G. and B. Marangoni. *"Hormonal changes during the rooting of hardwood cuttings of grapevine rootstocks"*. American Journal of Enology and Viticulture, Vol. 32, No. 2, pp. 135-137. (1981).

[12] Lee, I. J. *"Practical application of plant growth regulator on horticultural crops"*. J. Hort. Sci, Vol. 10, pp. 211-217. (2003).

[13] Ozgen, M. Serce, S. and Kaya, C. *"Phytochemical and antioxidant properties of anthocyaninrich Morus nigra and Morus rubra fruits"*. Sci. Horti., Vol. 119, No. 3, pp. 275-279. (2009).

[14] Parvez, M. Zubair, M. Saleem, M. Wali, K. and M. Shah. "*Effect of Indolebutyric Acid (IBA) and planting times on the growth and rooting of Peach cuttings*". Sarhad Journal of Agriculture, Vol.23, No.3, pp. 587-592. (2007).

[15] Phipps, H.M., Hansen, E.A. and A.S. Fege. *"Preplant soaking of dormant Populus hardwood cuttings [Vegetative propagation, artificial regeneration]"*. USDA Forest Service research paper NC-United States. https://www.nrs.fs.fed.us/pubs/rp/rp_nc241.pdf. (1983).

[16] Polat, A.A. *"Effect of Indolebutyric Acid on Rooting of Mulberry Cuttings"*. In XXVII International Horticultural Congress-IHC2006: International Symposium on Endogenous and Exogenous Plant Bioregulators, Vol.* 774, pp. 351-354. (2006).

[17] Ruter, J.M. *"Cloning Plants by Rooting Stem Cuttings"*. In Beyl, C. A., & Trigiano, R. N. (eds.) Plant propagation concepts and laboratory exercises. CRC Press, pp. 219-259. (2014).

[18] San José, M. C., Romero, L. and L.V. Janeiro. *"Effect of indole-3-butyric acid on root formation in Alnus glutinosa microcuttings"*. Silva Fenn, Vol, 46 No. 5, pp. 643-654. (2012).

[19] Schaff, S. D., Pezeshki, S. R., and F.D. Shields Jr. *"Effects of Pre Planting Soaking on Growth and Survival of Black Willow Cuttings"*. Restoration Ecology, Vol., 10, No. 2, pp. 267-274. (2002).

[20] Van der Krieken, W. M., Breteler, H., Visser, M. H. and D. Mavridou. *"The role of the conversion of IBA into IAA on root regeneration in apple: introduction of a test system"*. Plant Cell Reports, Vol. 12, No. 4, pp. 203-206. (1993).

[21] Wiesman, Z., Riov, J. and E. Epstein. *"Characterization and rooting ability of indole-3-butyric acid conjugates formed during rooting of mung bean cuttings"*. Plant physiology, Vol. 91, No. 3, pp. 1080-1084. (1989).

[22] Woodward, A.W. and B. Bartel. *"Auxin: regulation, action, and interaction"*. Annals of botany, Vol. 95 No. 5, pp. 707-735. (2005).